\documentclass[aps,twocolumn,groupedaddress,superscriptaddress,floatfix,amsmath,amssymb,prb]{revtex4-1}

\usepackage[version=3]{mhchem}
\usepackage[normalem]{ulem}

\usepackage{fancyhdr}
\usepackage{kantlipsum}
\usepackage{upgreek}
\usepackage{graphicx}
\usepackage{stmaryrd}
\usepackage{color}
\usepackage{siunitx}
\usepackage{hyperref} 
\usepackage{breakurl}
\usepackage{lettrine}

\newcommand{\nanotec}{
CNR NANOTEC − Institute of Nanotechnology, via Monteroni, 73100, Lecce, Italy}
\newcommand{\unisal}{
Dipartimento di Matematica e Fisica, Universit\`{a} del Salento, Via Arnesano, 73100 Lecce, Italy
}
\newcommand{\pavia}{
Dipartimento di Fisica, Universit\`{a} degli Studi di Pavia, Via Bassi 6, 27100 Pavia, Italy}
\newcommand{\brasil}{
Universidade Federal de Minas Gerais, Avenida Presidente  Antonio Carlos, 6627 - 31270901, Belo Horizonte, Brazil}
\newcommand{\ferrara}{
Department of Physics and Earth Sciences, University of Ferrara, Via G. Saragat 1,  I-44122 Ferrara, Italy}
\newcommand{\infn}{
INFN Istituto Nazionale di Fisica Nucleare, Sezione di Lecce, 73100 Lecce, Italy}
\begin{document}

\title{ Tunable out-of-plane excitons in 2D single crystal perovskites }

\author{A.~Fieramosca}
\affiliation{\nanotec}
\affiliation{\unisal}

\author{L.~De~Marco}
\email{luisa.demarco@nanotec.cnr.it}
\affiliation{\nanotec}
\author{M.~Passoni}
\affiliation{\pavia}
\author{L.~Polimeno}
\affiliation{\nanotec}
\affiliation{\unisal}
\author{A.~Rizzo}
\affiliation{\nanotec}
\author{B.~L.~T.~Rosa}
\affiliation{\brasil}
\author{G.~Cruciani}
\affiliation{\ferrara}
\author{L.~Dominici}
\affiliation{\nanotec}
\author{M.~De~Giorgi}
\affiliation{\nanotec}
\author{G.~Gigli}
\affiliation{\nanotec}
\affiliation{\unisal}
\author{L.~C.~Andreani}
\affiliation{\pavia}
\author{D.~Gerace}
\email{dario.gerace@unipv.it}
\affiliation{\pavia}
\affiliation{\nanotec}
\author{D.~Ballarini}
\email{dario.ballarini@nanotec.cnr.it}
\affiliation{\nanotec}
\author{D.~Sanvitto}
\affiliation{\nanotec}
\affiliation{\infn}

\begin{abstract}
Hybrid organic-inorganic perovskites have emerged as very promising materials for photonic applications, thanks to the great synthetic versatility that allows to tune their optical properties. 
In the two-dimensional (2D) crystalline form, these materials behave as multiple quantum-well heterostructures with stable excitonic resonances up to room temperature. 
In this work strong light-matter coupling in 2D perovskite single-crystal flakes is observed, and the polarization-dependent exciton-polariton response is used to disclose new excitonic features. 
For the first time, an out-of-plane component of the excitons is observed, unexpected for such 2D systems and completely absent in other layered materials, such as transition-metal dichalcogenides. 
By comparing different hybrid perovskites with the same inorganic layer but different organic interlayers, it is shown how the nature of the organic ligands controllably affects the out-of-plane exciton-photon coupling. 
Such vertical dipole coupling is particularly sought in those systems, e.g. plasmonic nanocavities, in which the direction of the field is usually orthogonal to the material sheet.
Organic interlayers are shown to affect also the strong birefringence associated to the layered structure, which is exploited in this work to completely rotate the linear polarization degree in only few microns of propagation, akin to what happens in metamaterials. 
\end{abstract}

\maketitle

\vspace{0.3cm}


\section*{Introduction }

\noindent 
Two-dimensional (2D) semiconductors attract increasing attention, due to both their relevance in quantum optoelectronics and their complexity in the physical properties, still not fully unraveled. 
In addition to many single-layer systems, such as transition-metal dichalcogenides (TMD), graphene and its inorganic analogues, hybrid 2D organic-inorganic perovskites offer a valuable alternative because they may be represented as natural realizations of multiple quantum-well (QW) heterostructures, with outstanding optical properties at room temperature~\cite{Pedesseau2016, Saparov2016,Xiong2017,Brandon2016, soci2017}.
Layered 2D perovskites generally consist of an inorganic layers of [PbX\textsubscript6]$^{2-}$ octahedra (with the halogen X$=$Cl, Br or I) sandwiched between bilayers of intercalated alkylammonium cations (see, e.g., {Figure~1}a). 
The lowest-energy electronic excitations are associated to the inorganic sheet, while the organic part is believed to behave as a potential barrier~\cite{Saparov2016}. Therefore, these crystalline materials combine the advantages of organics, such as the easy and cheap manufacturing, and those possessed by inorganic compounds, i.e. robustness and excellent optical properties~\cite{soci2017metasurface}. Moreover, by changing the inorganic precursors it is possible to tune the bandgap on a wide energy range, while the choice of the organic component can tailor the QW-type structure to a large extent. As compared to epitaxially-grown GaAs-based QW heterostructures, for instance, these materials allow for higher dielectric confinement and larger binding energies. As a further advantage over TMD monolayers, 2D hybrid perovskites display enhanced collective effects due to the large number of layers stacked in a single crystal without the problem of moving towards an indirect band gap typical of bulk TMDs. 

These peculiar properties make layered 2D perovskites not only an interesting system to be investigated per se but also ideal candidates for the development of novel optoelectronic devices to efficiently control photonic signals~\cite{RevModPhys.85.299,Byrnes2014,Sanvitto2016}. Particularly interesting is the use of 2D perovskites as active layers for strong light matter coupling with the consequent generation of exciton-polariton quasiparticles~\cite{PhysRevLett.69.3314, Snoke2017, Mazzeo2014, delvalle2009, Dominici2015}.
Indeed, evidence for exciton-polariton effects has been reported, although only in polycrystalline 2D hybrid perovskite thin films embedded in mirror microcavities~\cite{PhysRevB.57.12428,Brehier2006,PhysRevB.74.235212,Lanty2008,Lanty2008b,niu_image_2015}, and recently in all-inorganic perovskite CsPbCl$_{3}$ nanoplatelets optically confined between two distributed Bragg reflectors~\cite{Su2017}.

Here, we observe strong exciton-photon coupling in 2D hybrid perovskites single-crystals waveguides avoiding the need to embed them within mirror microcavities and we use polarization-dependent exciton-polariton response to investigate their excitonic properties. Taking advantage of their strong light-matter coupling, we assess the nature of the elementary excitations of these systems and varying the polarization of the incident electromagnetic radiation we probe the in-plane and out-of-plane excitonic response. Surprisingly, for the first time we disclose a polariton response associated to an out-of-plane excitonic component which was somehow unexpected in such 2D structures.
This observation was possible thanks to the use of large single crystal flakes of 2D perovskite that are employed to directly measure the excitonic response without being limited by non-radiative losses and grain-to-grain heterogeneity or tilted boundaries, usually present in polycrystalline films. \\
The dual nature of 2D perovskite excitons is also independently confirmed by far field polarized photoluminescence measurements which allow us to assess the in-plane and out-of-plane components of the exciton transition momentum dipole.
Moreover, to investigate the origin of this behavior we produce 2D layered crystalline perovskites with the same inorganic part but different organic ligands, and we observe that the dual-exciton features are related to the inter-QW distance and layer structure, 
 showing that the choice of the organic component provides a new route to control not only the energy but also the orientation and polarization of the transition dipole in these hybrid perovskites.

Finally, we measure the optical anisotropy due to the layered electronic structure akin to a metamaterial, finding that the choice of organic interlayer sensitively alters the polarized optical response. 
We observe large intrinsic birefringence in our mono-crystalline flakes and we demonstrate the use of them as half-waveplates with thicknesses of few optical periods.

\section*{Results }

We produce large thin crystals of 2D hybrid perovskite by an Anti-solvent Vapor assisted Crystallization method recently reported~\cite{Ledee2017} and subsequent mechanical exfoliation.
We synthesize three different layered perovskites changing the organic precursors: we select two alkyl cations, butylammonium (C$_{4}$H$_{9}$NH$_{3}$)$_{2}$PbI$_{4}$ (BAI), and octylammonium (C$_{8}$H$_{17}$NH$_{3}$)$_{2}$PbI$_{4}$ (OCT), having different chain lengths in order to tune the distance between the inorganic layers; in addition we introduce an aromatic moiety, phenethylammonium (C$_{6}$H$_{5}$(CH$_{2}$)$_{2}$NH$_{3}$)$_{2}$PbI$_{4}$ (PEAI), to investigate its effect on the exciton-photon coupling. Further details about their synthesis and their structures are provided in the Appendix. 

\begin{figure}[ht]
\centering
\includegraphics[width=0.48\textwidth]{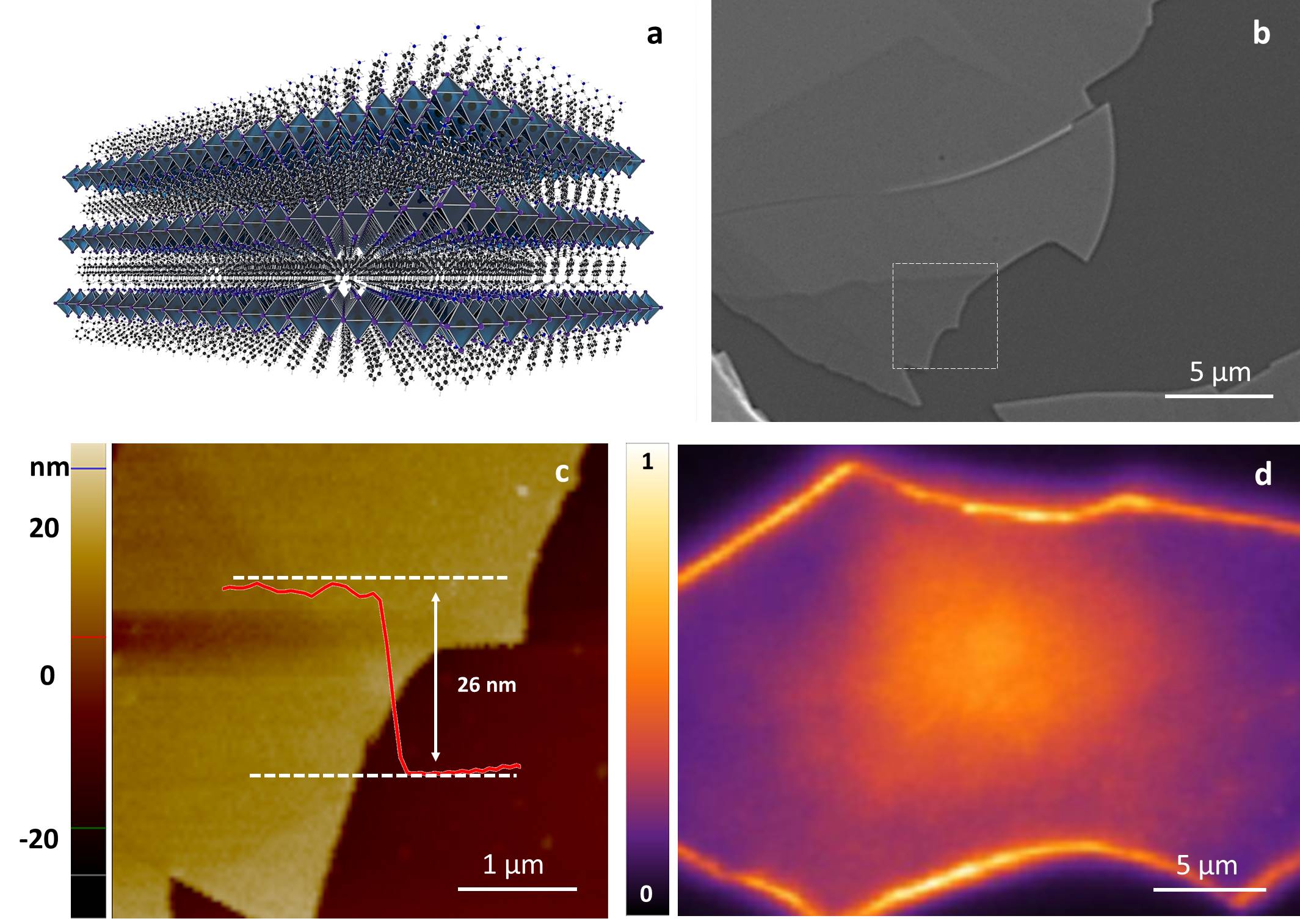}
\caption{a) Schematic illustration of the layered perovskite crystal structures; b) SEM image of a (PEAI)\textsubscript2PbI\textsubscript4  exfoliated crystal; c) AFM topography image of the white dashed square indicated in b with the corresponding height profile;   d)Real space photoluminescence of a (PEAI)\textsubscript2PbI\textsubscript4 perovskite crystal.   }\label{fig:1} 
\end{figure}

\begin{table}[hb!]
  \includegraphics[width=0.5\textwidth]{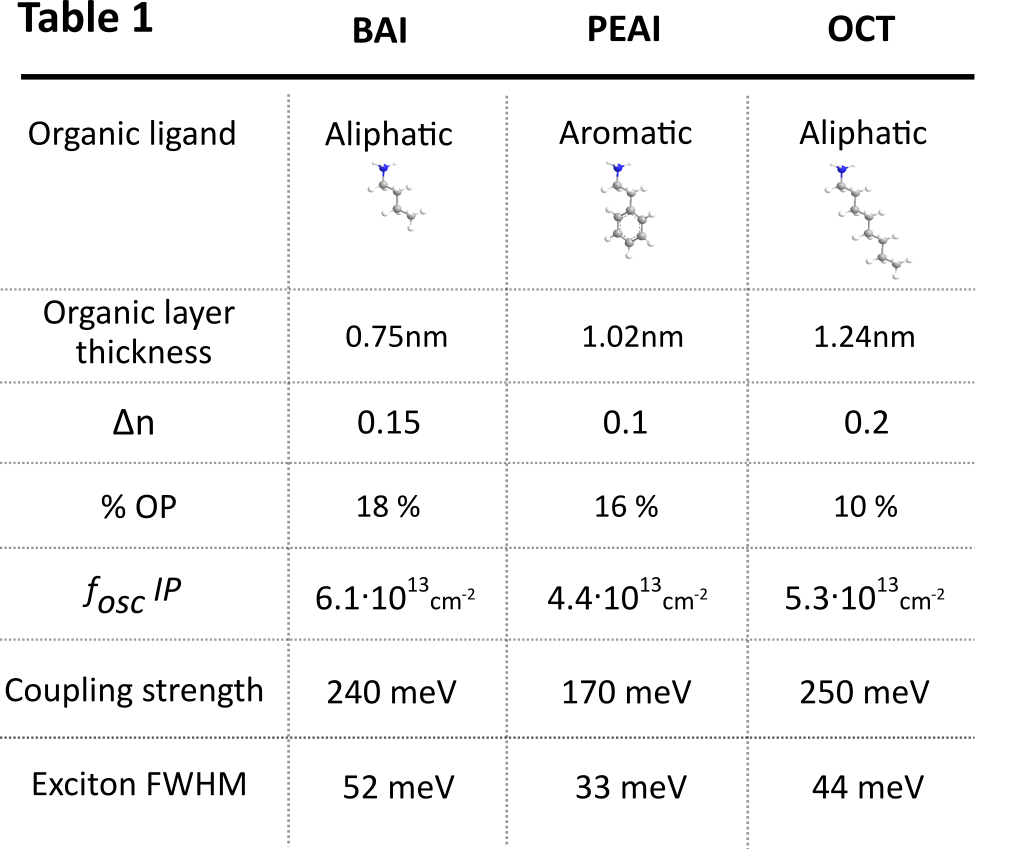}
  \caption{Parameters obtained from the comparison between numerical simulations and experimental data, for the three different types of layered 2D perovskite films considered.}
\end{table}

The unit cell parameters verified by X-ray diffraction measurements (see Figure~5 and Table~II in the Appendix) allow to determine the thicknesses of barrier and well in the multi-QW structure: the inorganic semiconductor layer is 0.64~nm thick while the organic interlayer is 0.75~nm, 1.02~nm and 1.24~nm thick for BAI, PEAI and OCT, respectively (see {Table~I} and Table~II).
The exfoliated flakes observed by scanning electron microscopy (SEM) and atomic force microscopy (AFM), shown in Figure~1b, Figure~1c (see also Figure~6 and Figure~7 in the Appendix), reveal terrace structures with the planes oriented parallel to the substrate. 
All samples show an extremely uniform photoluminescence (see Figure~1d) that, differently from the majority of the reported studies carried out on polycrystalline films, is not affected by the presence of inter-grains voids or grain boundaries (see also Figure~8 in the Appendix). In a single crystal we can notice that the bright regions in the PL emission observed along the boundaries of the flake are due to scattering of the guided radiation with defects at the edges of the film, while the rest of the crystal is very uniform and defect free including the excitation spot which is rather brighter in the central part of the flake. Absorption and photoluminescence (PL) spectra taken on thin single-crystal flakes are shown in the Appendix (Figure~9). 

\begin{figure*}[ht]
\centering
\includegraphics[width=0.8\textwidth]{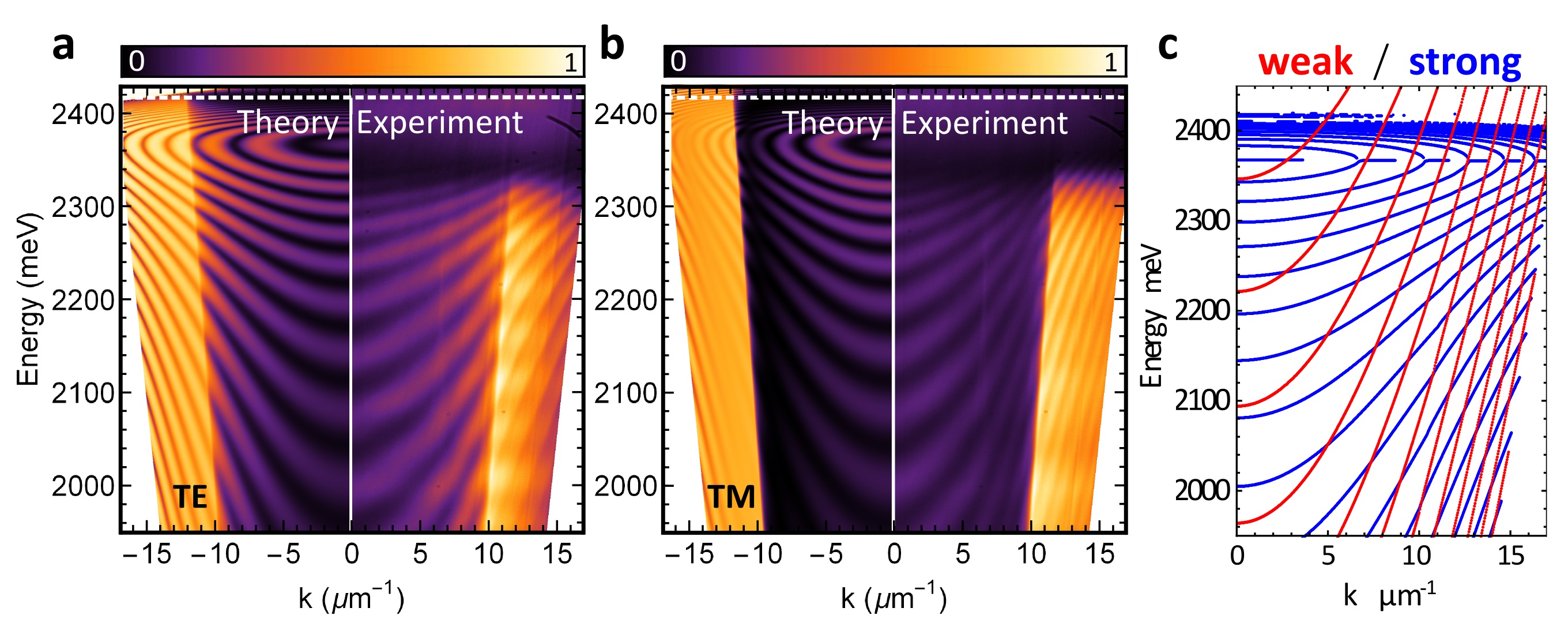}
\caption{Reflection spectra are plotted as energy versus in-plane momentum, $k$, with $k=\frac{2 \pi}{\lambda}\sin{\theta}$, where $\lambda$ is the wavelength and $\theta$ the incidence angle. Multiple resonances are due to the Fabry-Perot configuration, with the enhanced intensity corresponding to angles of incidence beyond the air light line. Reflectivity maps measured (right half-panel) and calculated by scattering matrix (left half-panel) for OCT single crystal under white light illumination for (a) TE and (b) TM polarizations are shown in left and right column, respectively. (c) Reflection minima calculated with (blue lines) and without (red lines) the excitonic resonance, showing the crucial modification of the reflection spectrum induced by strong light-matter interaction as compared to a bare film response.
}\label{fig2}
\end{figure*}

The optical response of single crystals of BAI, PEAI and OCT with thicknesses of few microns is investigated with an oil-immersion objective (details are reported in the Appendix) that allows to capture, from the glass substrate side, also the signal beyond the air light-line, i.e. the signal coming from total internal reflection (TIR) at the crystal-air interface. Both PL and reflection spectra (see {Figure~2}a,b and also Figs.~10 and 11 in the Appendix) show the frequency modulation due to partial reflection on the crystal-substrate (glass) interface and to TIR at the crystal-air interface for angles larger than $\theta=\arcsin{\frac{n_{air}}{n_{glass}}}\approx40^{\circ}$, where $n_{air/glass}$ is the refractive index of vacuum (or substrate, respectively). The figure shows the typical dispersion of strong light-matter coupling, with the optical resonances bending close to the excitonic resonance at large angles.
A considerable amount of information about the nature of the elementary excitations in these compounds can be inferred by analyzing the material response to optically exciting polarizations with either the electric field (TE, Figure~2a) or the magnetic field (TM, Figure~2b) transversely oriented with respect to the incidence plane. Indeed, for TE polarized light, only the dipoles oriented in the plane of the QW (associated to the [PbX\textsubscript6]$^{2-}$ inorganic sheets) contribute to the material polarizability, while both in-plane (IP) and out-of-plane (OP) dipole strengths have to be considered when looking at the response in TM polarization. By comparing the experimental data (right-side of each panel in Figure~2) with numerical simulations (left-side of each panel in Figure~2), we are able to obtain a quantitative estimation on the exciton dipole orientation in these materials. 

The optical response of the 2D perovskite films is modeled with a generalized scattering matrix algorithm, ~\cite{Li1996,Whittaker1999,Liscidini2008} which allows to simultaneously include both the natural birefringence of the stacked organic-inorganic layers and the polarization-dependent exciton response. Details about the theoretical method and the implementation used in this work are reported in the Appendix. Here we stress that this modeling is essential to capture the key ingredients of such peculiar systems, which would not be amenable to be simulated with standard transfer matrix approaches~\cite{Andreani1994,Vladimirova1996,Lerario2014}. Thanks to this analysis we have selected the values that allow us to best describe the qualitative and quantitative behavior of the different layered materials, as summarized in Table~1. Doing so we can demonstrate that the optical response of these materials is determined from doubly polarized excitonic contributions, corresponding to optically active dipoles oscillating both in and out of the inorganic layers plane. This is surprising, since these materials are considered 2D semiconductors with excitons strongly confined between insulating barriers. Nevertheless, the vertical contribution is crucial to correctly capture the reflectivity spectra detected in TM polarization 
with out-of-plane (OP) oscillating dipoles, contributing significantly to the oscillator strength of the radiation-matter coupling, although by a generally smaller amount as compared to the in-plane (IP) oscillating dipoles (see quantitative estimate in Table~I).
In addition, we observe that the out-of-plane oscillator strength decreases on increasing the organic barrier thickness between the inorganic perovskite layers (see the values obtained for BAI, PEAI, and OCT in  Table~I): longer organic chains seems to lead to a stronger confinement in the QW plane, with the exception of benzene rings that slightly enhance the out-of-plane polarizability and strongly reduce the optical birefringence with respect to aliphatic chains of similar length. 

\begin{figure}[ht]
\centering
\includegraphics[width=0.48\textwidth]{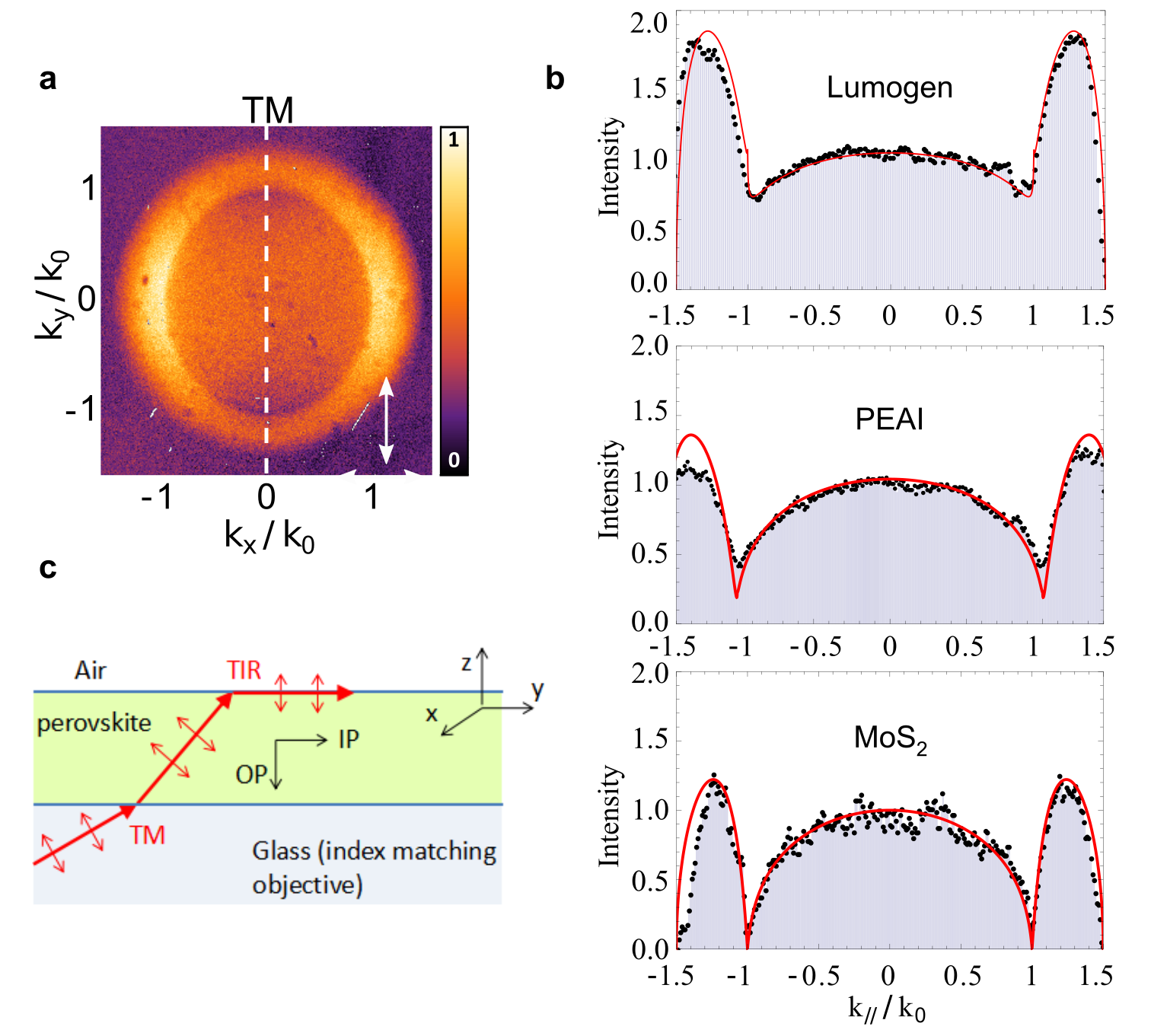}
\caption{a) 2D Fourier space emission (PL) in the linear vertical polarization (white arrow). The white dashed line represent the TM section, in which the OP contribution is distinguished at the TIR angle (corresponding to $k_{0}$ for each material); b) PL signal along the white dashed line in a) for isotropic material (Lumogen), thin crystal 2D perovskite (PEAI) and MoS\textsubscript{2} monolayer. c) Scheme representing the vanishing local optical density of states at $k_{\parallel}/k_{0}=1$ for out-of-plane dipoles in vertical polarization. In the TM plane, the OP fraction is 50\%, 30\% and 0\% for Lumogen, PEAI and MoS\textsubscript{2}, respectively. These values correspond, in the 3D volume, to an OP contribution of 33\% for the isotropic case and 18\% for the PEAI.}
\end{figure}

To further corroborate the quantitative estimate of an out-of-plane component to the exciton-polariton response we use Fourier-resolved, polarized photoluminescence measurements on a thin PEAI single crystal of 20~nm. 
This is an effective technique to obtain the IP and OP contributions for thin films~\cite{schuller_orientation_2013,Scott2017}. The Fourier plane of the vertical polarization is imaged in {Figure~3}a. For the emission direction along $k_{y}$, the local density of optical states is completely suppressed close to the TIR direction for IP polarization, and the OP contribution can be isolated. In Figure~3b, the results for the PEAI crystal are compared with a MoS\textsubscript{2} monolayer (fully in-plane excitons) and with an isotropic molecular thin film (Lumogen Orange, 35 nm). The fitting functions are obtained by following the analytic model in Ref~\onlinecite{schuller_orientation_2013,Scott2017}, and using an OP fraction of $(33\pm5)\%$  for the isotropic case and $<5\%$ for MoS\textsubscript{2} monolayer. According to the results of simulations in Figure~2, we consider an OP fraction of $(18\pm5)\%$ for the PEAI thin film, obtaining an overall good agreement with both the experimental results and the scattering matrix analysis. This measurement independently confirms the presence of out-of-plane oscillating dipoles and thus the double nature of 2D perovskite excitons.

These results show two particularly relevant aspects: on the one hand, we clearly evidence a polarization-dependent and anisotropic (i.e., possessing an OP dipolar component) exciton response, on the other hand we see a barrier-dependent oscillator strength. 
Both these effects are unexpected at first sight, since up to now electronic band structure calculations have predicted an ideally perfect in-plane orientation of the exciton momentum dipole due to the high energy barrier ($\SI{\sim1}{\electronvolt}$) of the organic layer. This evidences that the role of the organic ligands had been downplayed in these compounds. ~\cite{Even2012}

As a matter of facts it is well estabilished that the elementary excitations in these layered materials should arise from optical transitions coming from half-integer total angular momentum valence band states ($j=1/2, m_j = \pm 1/2$, of Pb-6s orbitals) into conduction band states with the same $j,m_j$ symmetry (mainly arising from Pb-6p orbitals, see e.g. Ref.~\onlinecite{Giovannie1600477,Yu2016, Umebayashi2003}). This is quite different to what happens in inorganic QWs, i.e. III-V semiconductors, in which heavy-hole excitons represent the lowest energy transitions from $j=3/2$ (p-type) valence band to $j=1/2$ (s-type) conduction band states. While these latter transitions leads to strictly in-plane polarized states (see, e.g., Ref.~\onlinecite{cardonabook}), in the case of 2D perovskites -- due to polarization selection rules derived from dipole matrix elements -- vertically polarized transitions are permitted.

Our observations suggest that the organic interlayer could actually play a non-negligible role beyond that of a passive insulating barrier. Moreover, the dielectric constant of the organic part, its thickness and its energy levels could affect the interactions between the excitons of the semiconducting inorganic layers, and thus their response to a given polarization of the incident radiation~\cite{Pedesseau2016}.
These hypotheses are only speculative and worth being investigated in detail, e.g. through microscopic theories of the optical response that the present work will hopefully stimulate. 

Regarding the interest of these results for polaritonic applications, we observe that the values of the exciton oscillator strength per unit surface (reported in Table~I) are fully consistent with the strong coupling regime in all of the perovskite samples considered so far. In particular, the radiation-matter coupling energy can be obtained by the usual expression employed for bulk exciton-polaritons, which is expressed as~\cite{gerace2007}
\begin{equation}
E_c= \hbar \Omega_c= \hbar \sqrt{\frac{e^2}{4\varepsilon_0 \varepsilon_r m_0}\frac{f}{V}}
\end{equation} 
where $e$ is the elementary electric charge, $m_0$ is the free electron mass, and $\varepsilon_r$ is the background dielectric permittivity relative to the vacuum, $\varepsilon_0$. 
Here, the 2D nature of the elementary excitations in each perovskite layer is combined with the multi-QW type structure, which is interpreted as an effective medium with oscillator strength per unit volume $f/V=(f/S)/d$, where $d$ is the thickness of the active organic plus the inorganic barrier layers in each of the perovskite compounds considered. The estimated coupling strengths for the BAI, PEAI, and OCT materials (see Table~I) show values that are compatible with very large light-matter coupling regimes, in particular the condition $\hbar\Omega_{c}> FWHM$ is fully satisfied in all the three cases. 
We also notice that these values are especially remarkable when compared to standard III-V semiconductor heterostructures, in which radiation-matter coupling energy is typically measured in the order of few meV while the coupling strength is in the order of 200 meV in our case. 
In particular, the present experiments correspond in spirit to early investigations on the optical response of thin films in III-V semiconductors with strong excitonic absorption~\cite{Chen1991,Tredicucci1993,Boemare1995}, where exciton-polariton effects were first evidenced even in the absence of bottom and top mirrors.

\begin{figure}[ht]
\centering
\includegraphics[width=0.36\textwidth]{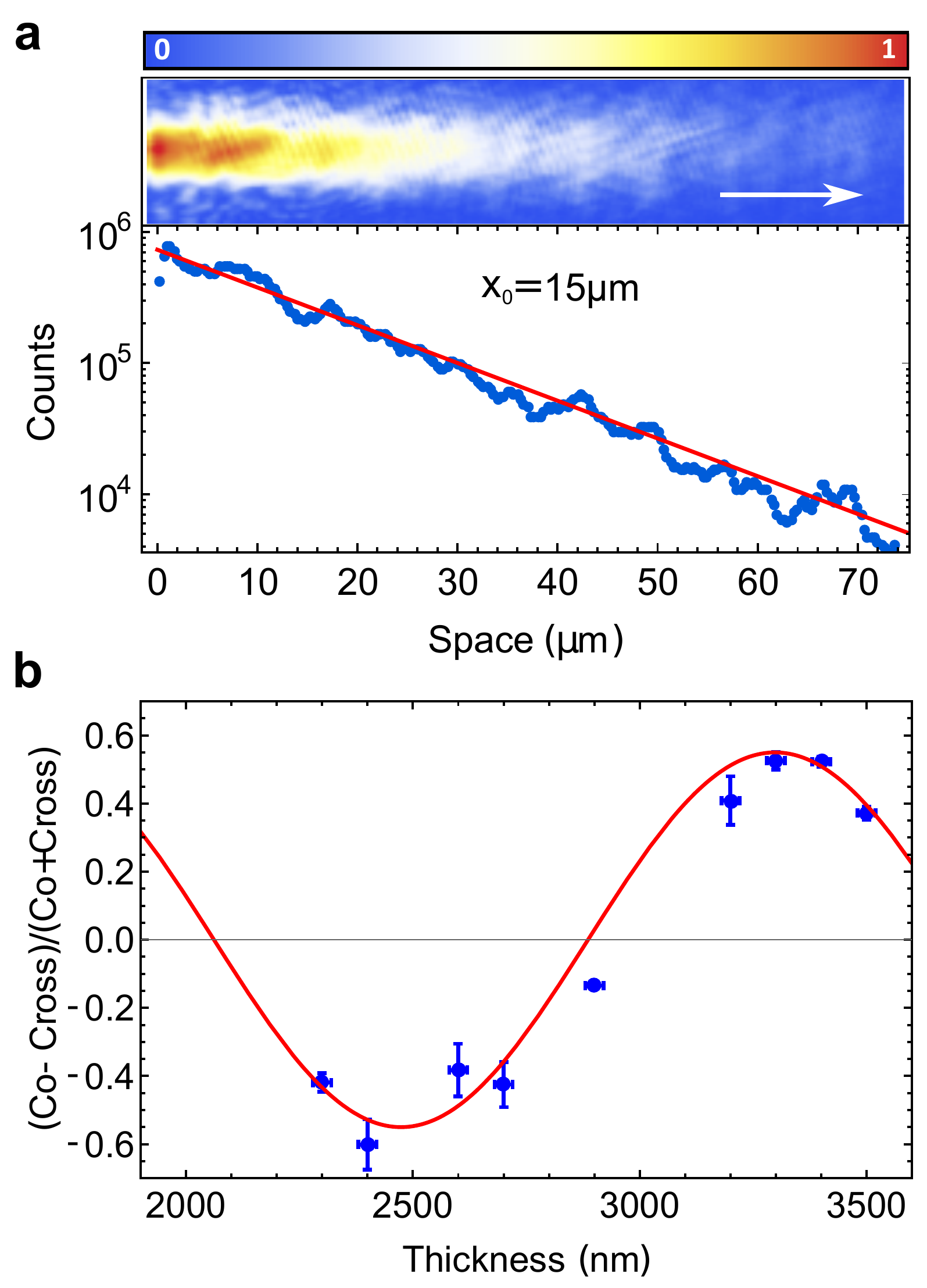}
\caption{a) Real space intensity image of a resonant injected beam in a PEAI crystal. b) Rotation of the linear polarization degree in a set of PEAI crystals with a birefringence of $\Delta n=0.1$. A continuous wave laser at 640 nm with diagonal polarization (mixed TE and TM), is focused by the oil-immersion objective beyond the angle of TIR at the air interface. The co- and cross-polarized reflected beam is measured for different slab thicknesses, and the linear polarization degree is fitted with a sinusoidal function.}
\end{figure}

Finally, we take advantage of the surface homogeneity of these monocrystalline slabs to illustrate optical propagation for distances of tens of microns without being scattered by defects (see {Figure~4}a and Figure~12 in the Appendix). An additional advantage of the large extension of the single crystals is the possibility to exploit the large optical birefringence induced by the layered material structure, akin to that of metamaterials. Interestingly, black phosphorus has been recently proposed for fabrication of atomically-thin optical waveplates thanks to its in-plane birefringence as high as $\Delta n\approx0.245$~\cite{He2017}. Here we measure the dispersion of the refractive index along and perpendicularly to the layer structure in the transparent region of the spectrum for PEAI, BAI and OCT (see Figure~14 in the Appendix), finding comparable values ($\Delta n\approx0.2$ for OCT, see Table~I). However, compared to the black phosphorus, the 2D perovskites show an out-of-plane birefringence, that can be tuned by the choice of the organic ligand either by changing the interlayer separation, or by exploiting the organic-inorganic nature of higher electronic levels in the presence of short interlayer distances or $\pi$-conjugated organic systems (see the effect of benzene rings on optical anisotropy in Table I). Here, as a demonstrator, we show that such a huge optical anisotropy can be effectively used as an ultrathin waveplate. For light traveling with a finite angle, with respect to the direction normal to the structure, the retardation between the components of the electromagnetic field oscillating in- and out-of-plane are large enough to result in a complete rotation of the linear polarization degree in few optical periods. In Figure~4b, the angle of incidence is chosen beyond the light-line in air to detect the totally reflected signal at the TIR interface, allowing to measure a complete rotation of the linear polarization degree for crystal thickness of just 1 micron. 
In inorganic microcavities, the optical and excitonic anisotropies are enhanced under strong coupling regime, giving rise to strong spin-orbit effective interactions for polaritons propagating in the plane of the structure. These results indicate that self-assembled waveguides of 2D layered perovskites can be a promising system to study the effects of the strong optical anisotropy on the spin dynamics of exciton-polaritons and can be used as integrated waveplates in optical circuits, as well as in polariton-based devices. 

\section*{Conclusions }

In summary, we have shown that 2D perovskites possess dual excitonic properties that can be tuned by the choice of the organic cations which in turn affect the exciton confinement and its optical response, both in terms of the out-of-plane component of the transition dipole moment and the intrinsic birefringence of the structure. The presence of an hybrid organic and inorganic structure in 2D perovskites provide an ideal platform for developing novel multifunctional materials in which the crystalline architectures can be synthetically fine-tuned in order to provide a huge range of semiconductors with different properties. These observations pave the way for the design and fabrication of all-optical devices based on 2D perovskites.

\acknowledgments
The authors acknowledge the ERC project ”ElecOpteR” grant number 780757. GG gratefully acknowledge the project PERSEO-“PERrovskite-based Solar cells: towards high Efficiency and lOng-term stability” (Bando PRIN 2015-Italian Ministry of University and Scientific Research (MIUR) Decreto Direttoriale 4 novembre 2015 n. 2488, project number 20155LECAJ). A.R. gratefully acknowledges SIR ``Two-Dimensional Colloidal Metal Dichalcogenides based Energy-Conversion Photovoltaics'' (2D ECO), project number RBSI14-FYVD. The authors acknowledge Sonia Carallo for technical support and AFM measurements. 
D.S., D.B., A.F. and L.D.M. are grateful to St\'ephane K\'ena-Cohen and F\'abio Barachati for fruitful discussions.  \\
\textbf{Author contributions.} A.F, L.P performed the optical measurements, and A.F., L.P., L.D.M., B.L.T.R. prepared the material and the samples. M.P, L.C.A., and D.G, performed the numerical simulations and the theoretical modeling and interpretation of data. G.C. performed X-ray measurements. L.D.,M.D.G.,A.R,G.G, contribute to the experimental realization and interpretation of the results. L.D.M., D.G, D.B. and D.S. supervised the work. \\
\textbf{Published paper} ACS Photonics, 2018, 5 (10), 4179 - 4185 \\
\textbf{DOI: 10.1021/acsphotonics.8b00984}


\appendix

\begin{figure*}[ht]
	\centering
	\includegraphics[width=0.7\textwidth]{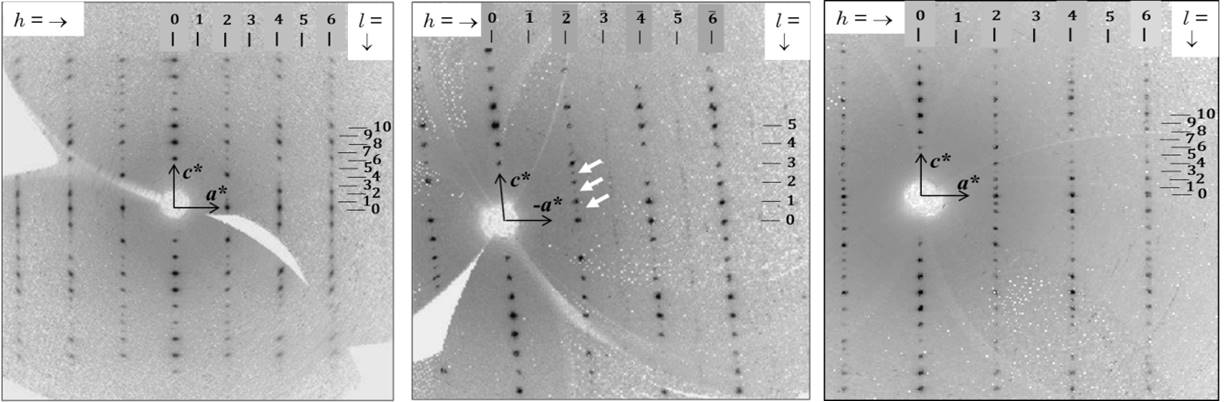}
	\caption{Precession images of the (a*-c*) 0-layer reciprocal plane for 2D-perovskites BAI (left), PEAI (middle), and OCT (right) as reconstructed from measured bidimensional single crystal X-ray diffraction frames. White arrows in PEAI mark the weak diffraction spots revealing a doubling of lattice periodicity along [001] (i.e. about 32.8 \r{A} instead of 16.4 \r{A}).}
\end{figure*}

\begin{figure*}[ht]
	\centering
	\includegraphics[width=0.7\textwidth]{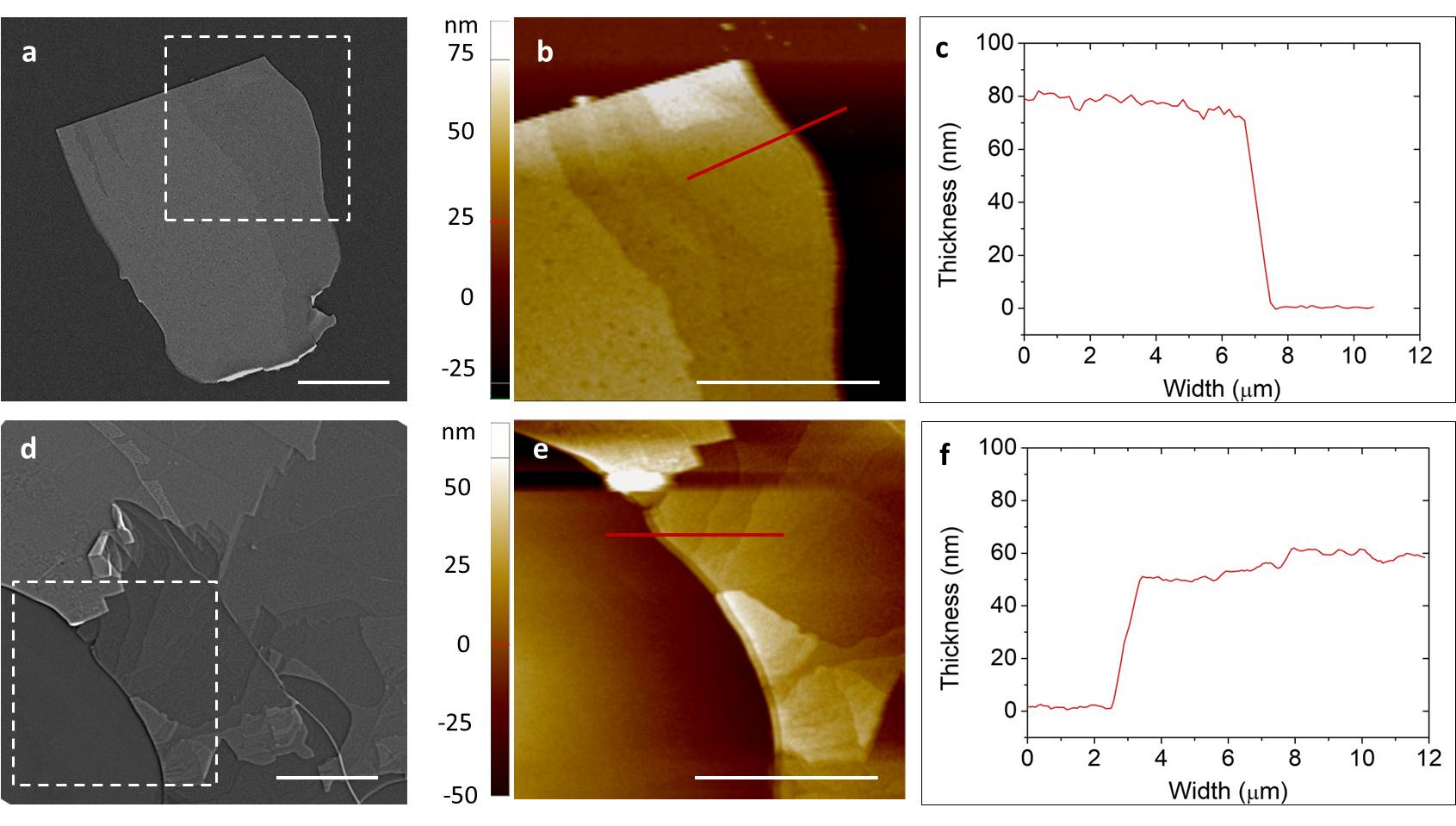}
	\caption{SEM images for (BAI)\textsubscript2PbI\textsubscript4 (a) and (OCT)\textsubscript2PbI\textsubscript4 (d) exfoliated crystals. AFM topography images of (BAI)\textsubscript2PbI\textsubscript4 (b) and (OCT)\textsubscript2PbI\textsubscript4 (e) in the white square indicated in (a) and (d), respectively. The corresponding height profiles are shown in (c) and (f). SEM scale bars: 20 micron; AFM scale bars: 10 micron}
\end{figure*}

\section{Experimental details}

\noindent{\bf Chemicals and Reagents}.
Phenethylammonium iodide and butylammonium iodide were purchased from Greatcell Solar. Octylamine, hydriodic acid (HI), dichloromethane, ethanol and diethyl ether were purchased from Sigma Aldrich. Lead(II) iodide (PbI\textsubscript2) was purchased from Alfa Aesar. Gammabutyrolactone was purchased from TCI. All chemicals were used as received without any further purification. SPV 224PR-M adhesive tape was kindly provided by Nitto Denko Corporation.

\noindent{\bf Synthesis of 2D perovskite flakes}.
Synthesis of Octylammonium Salt: 20 mL ethanol and 782 mg octylamine were added to a round-bottom flask to form a mixture which was kept at $ 0^\circ C$ using an ice bath. Then, a stoichiometric amount of HI (2.7 mL of concentrated 57 aqueous HI) was added dropwise. The mixture was stirred for 2.5 h to ensure fully reaction. The solvent was removed by rotary evaporator at $60^\circ C$ and the ammonium salt powder was collected and washed with diethyl ether for three times. Then the powder was dried at $60^\circ C$ under vacuum for 24 h. 

PEAI: 498 mg phenethylammonium iodide and 461 mg PbI\textsubscript2 were dissolved in 1 mL gammabutyrolactone and stirred at $70^\circ C $ for 1 hour.  
BAI: 403 mg phenethylammonium iodide and 461 mg PbI\textsubscript2 were dissolved in 1 mL gammabutyrolactone and stirred at $70^\circ C $ for 1 hour.  
OCT: 514 mg phenethylammonium iodide and 461 mg PbI\textsubscript2 were dissolved in 1 mL gammabutyrolactone and stirred at $ 70^\circ C $ for 1 hour. 

\begin{figure}[hb]
	\centering
	\includegraphics[width=0.48\textwidth]{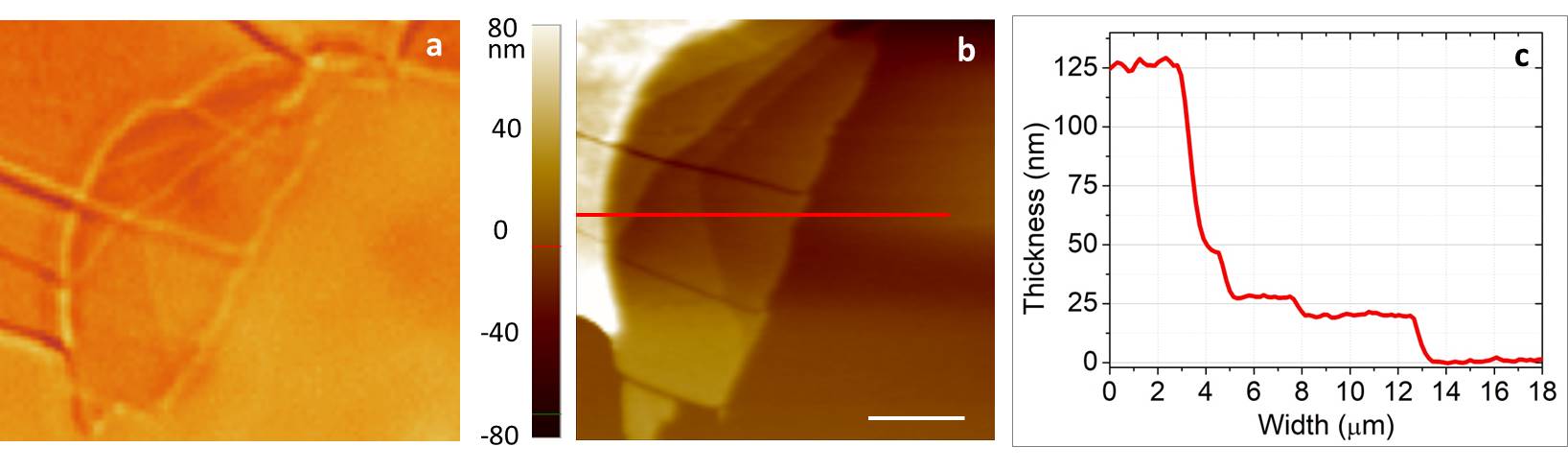}
\caption{a) Optical image of a PEAI single crystal taken in transmission mode; b) AFM topografy of the same crystal and c) corresponding height profile; AFM scale bar: 5 um.}
\label{PEAIthin}
\end{figure}

\begin{table*}[ht]
	\centering
	\includegraphics[width=0.7\textwidth]{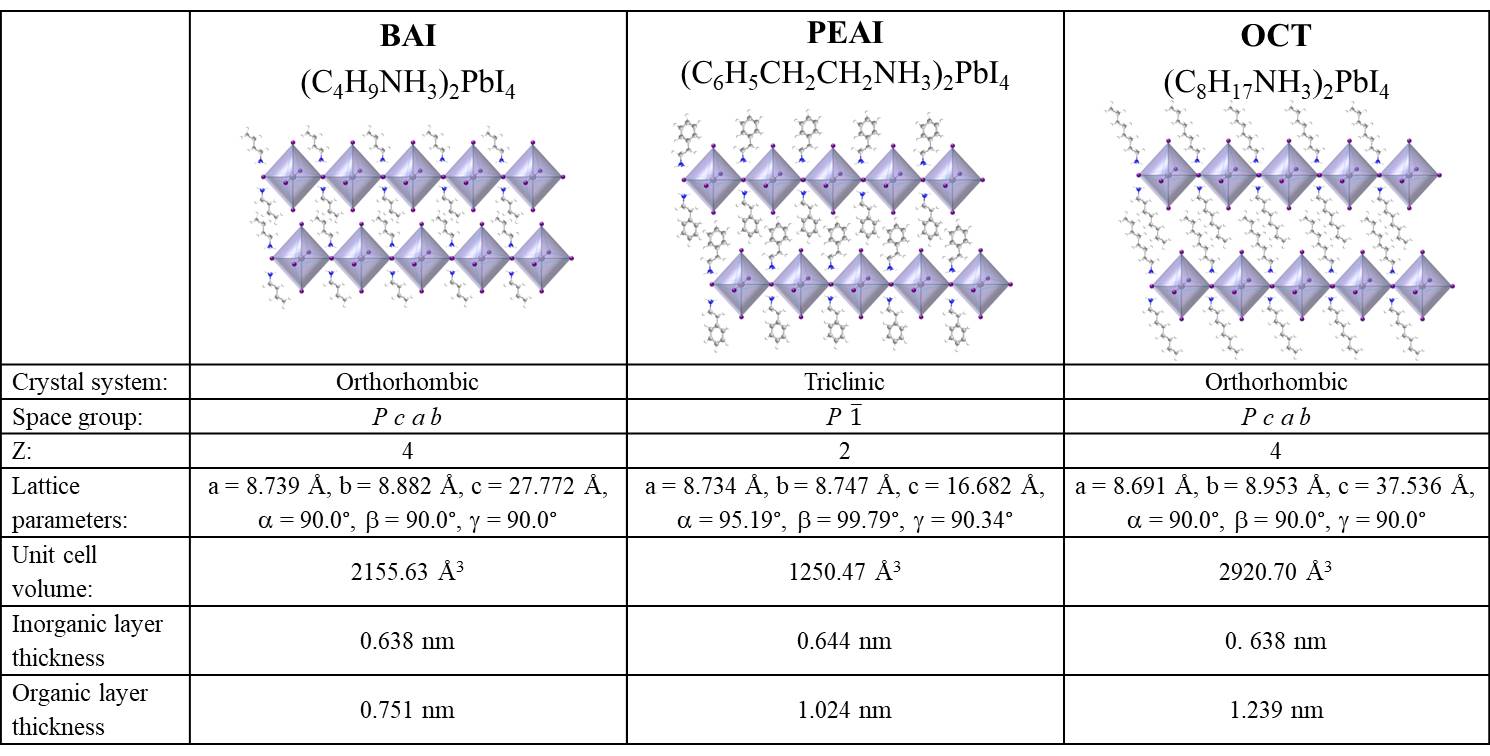}
	\caption{Summary unit cell parameters and crystal symmetry determined by single crystal XRD for (BAI)$_{2}$PbI$_{4}$, (PEAI)$_{2}$PbI$_{4}$ and (OCT)$_{2}$PbI$_{4}$; the thickness of the inorganic layers are determined by considering the Pb-I apical bond lengths while the thickness of the organic layer is given by the interlayer distance (calculated from the lattice parameters $c$) subtracted of the inorganic layer height.}
\end{table*}

\begin{figure}[hb]
	\centering
	\includegraphics[width=0.44\textwidth]{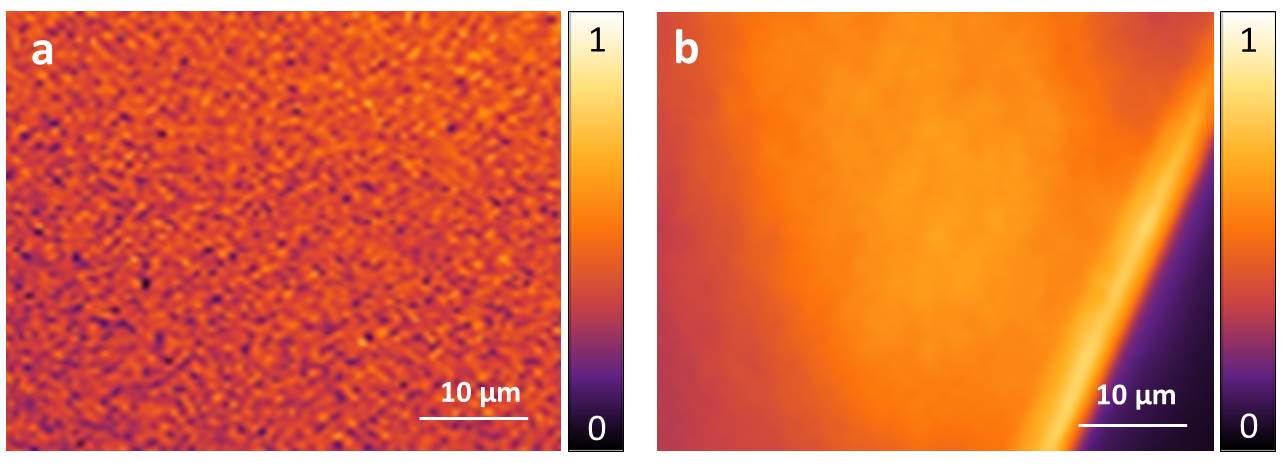}
	\caption{PL image of PEAI polycrystalline film (a) and PEAI single crystal flake (b).}
\end{figure}

\begin{figure}[hb]
	\centering
	\includegraphics[width=0.48\textwidth]{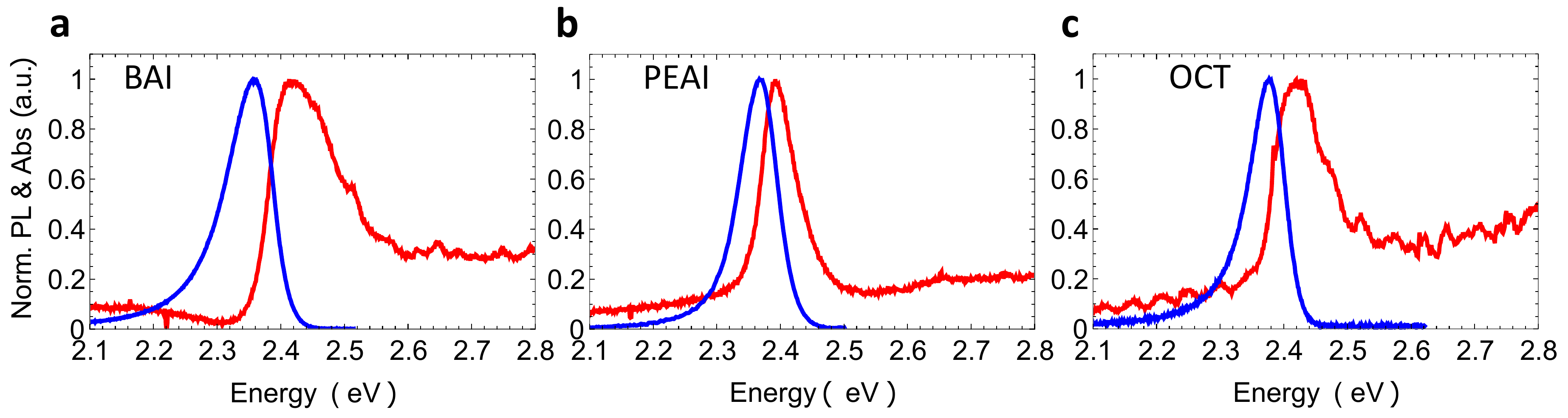}
	\caption{Absorption (Red line) and emission (Blue line) spectra for, respectively, BAI (a), PEAI (b) and OCT (c). A Xenon light source is used to measure the absorption while a CW laser (488nm) is used for PL measurements. The signal from the central part of the flake is considered.
}\label{figPL}
\end{figure}

2D perovskite single crystals were synthesized by Anti-solvent Vapor assisted Crystallization method as follows: 200 micron thick glasses were used as substrates. They were cleaned with acetone and water in ultrasonic bath for 10 min each and then soaked into a TL1 washing solution (H\textsubscript2O\textsubscript2/NH\textsubscript3/H\textsubscript2O 5:1:1, v/v), heated at $80^\circ C $  for 10 min to remove organic contamination and finally rinsed 10 times in water. 5 microliters of the perovskite solution are deposited on one of the substrate and immediately after capped by the second glass substrate. Then, a small vial containing 2 mL of dichloromethane is placed on the top of the two sandwiched substrates. Substrates and vial are placed in a bigger Teflon vial, closed with screw cap and left undisturbed for some hours. After this time millimetre-sized crystals appear in between the two substrates having a thickness varying from few to ten micrometres. In this approach, the perovskite precursors solution is exposed to a solvent in which the product is sparingly soluble (thus called anti-solvent); in this way supersaturation is easily reached and precipitation occurs since the solubility of 2D perovskites is drastically reduced. 
Single crystals are mechanically exfoliated with SPV 224PR-M Nitto Tape and transferred onto glass substrates. The exfoliated flakes, having the thickness of tens of nanometres, appear smooth and uniform over tens of square micrometers, as observed by scanning electron microscopy (SEM) and atomic force microscopy (AFM), see Fig.~6. Samples for SEM and AFM were exfoliated in the same manner and transferred onto Si substrate.

\begin{figure}[ht]
	\centering
	\includegraphics[width=0.48\textwidth]{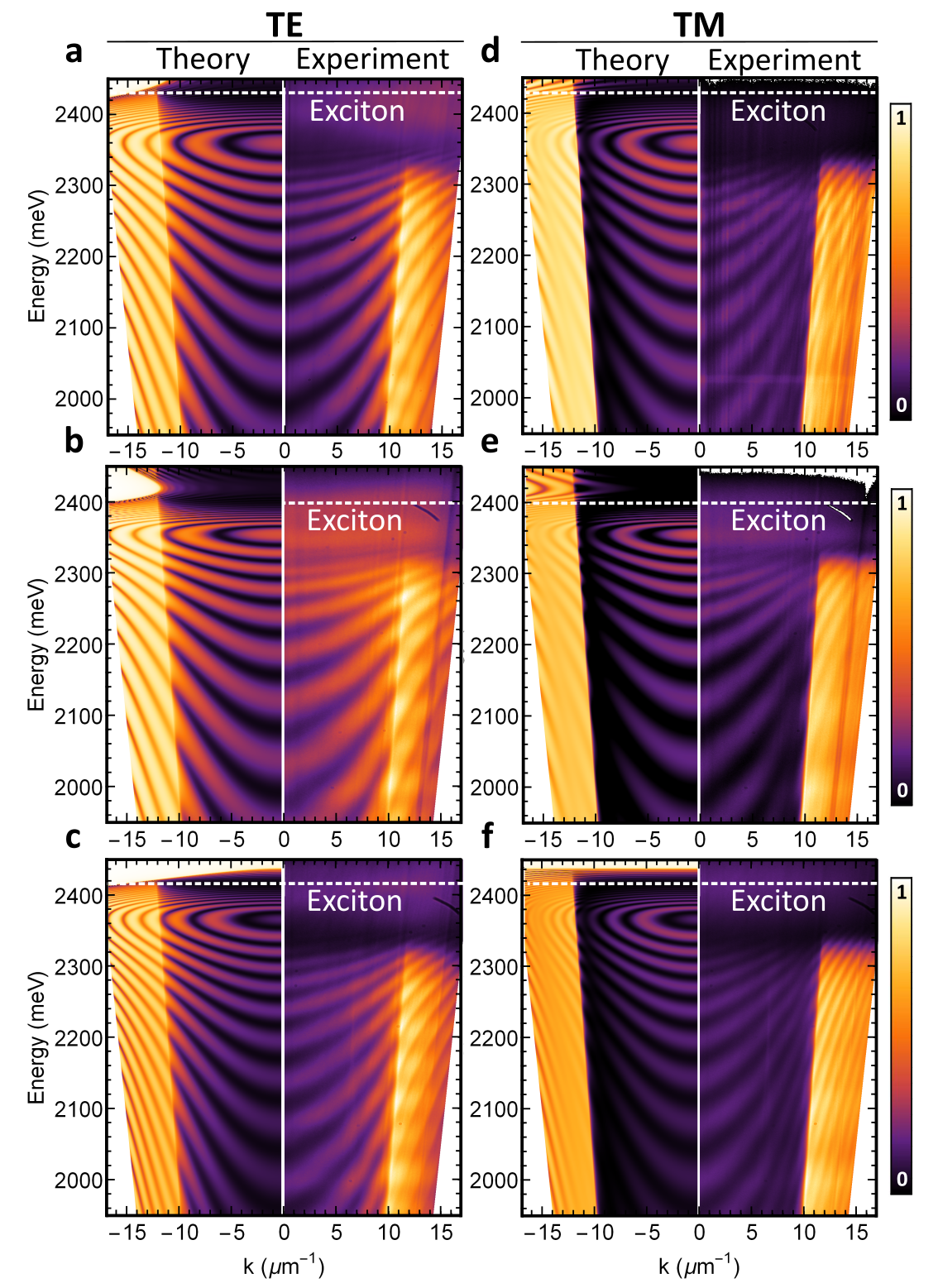}
	\caption{a), b), c) Energy-resolved reflection maps plotted as energy vs in-plane momentum k, measured (right) and calculated (left) for TE polarization for BAI, PEAI and OCT, respectively; d), e), f) Energy-resolved reflection maps, measured (right) and calculated (left) for TM polarization for BAI, PEAI and OCT, respectively.
}\label{fig:2}
\end{figure}

\begin{figure}[ht]
	\centering
	\includegraphics[width=0.48\textwidth]{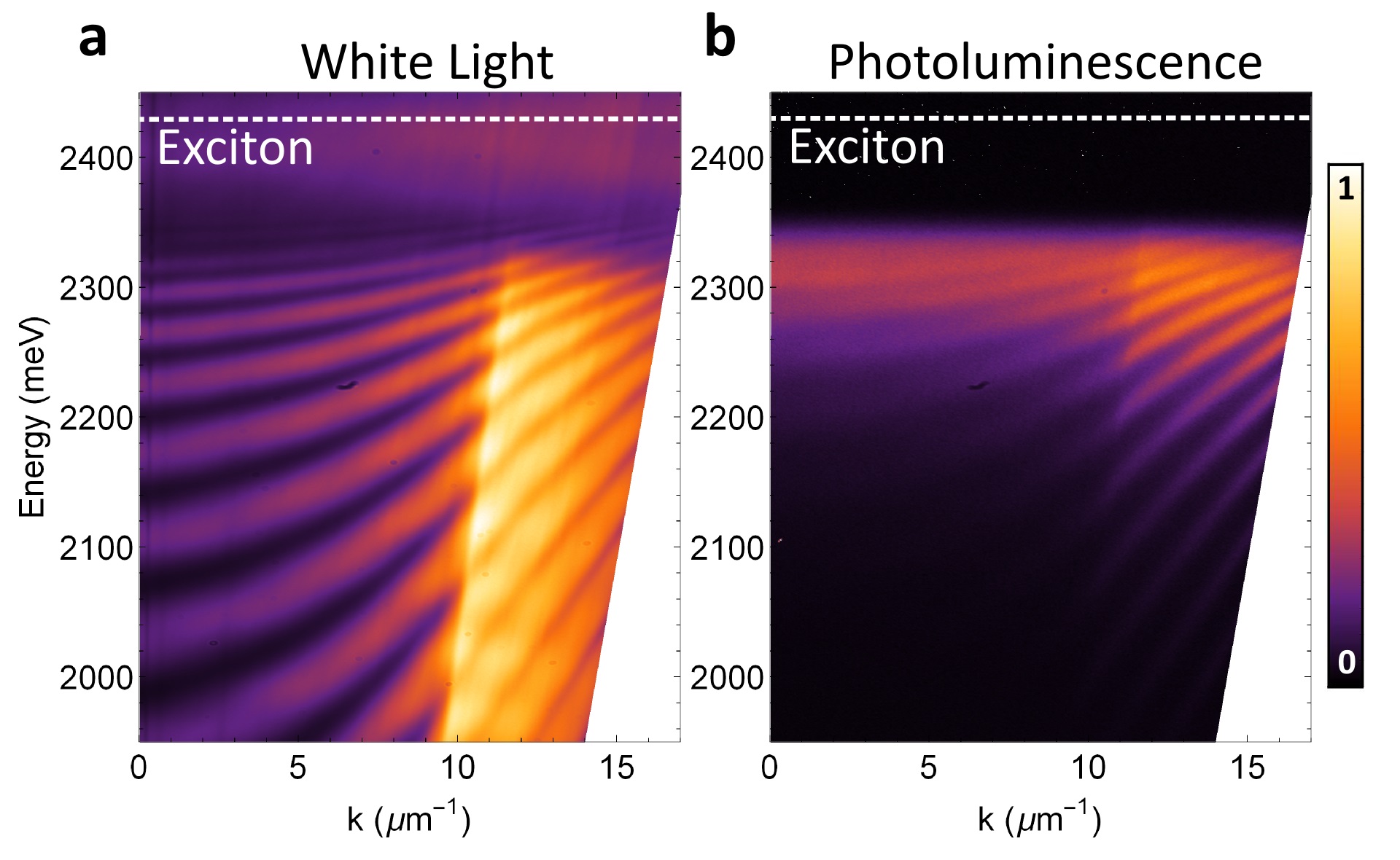}
	\caption{Energy vs in-plane momentum in a) reflection and b) emission configuration for TE polarization, taken in the same spatial position, for BAI crystal 2 um thick. Fabry-Perot resonances appear as dark minimum in reflection (a) and as bright maximum in PL (b).
}\label{fig:pulsed}
\end{figure}

\noindent{\bf Structural and morphological characterization}.
X-ray diffraction measurements were performed on single crystals of PEAI, BAI, and OCT 2D-perovskites picked from the glass slide. The smaller and larger dimensions of crystals were ranging from 7 to 15 um and from 65 to 73 um, respectively. Intensity data were collected using an automatic four-circle Nonius KappaCCD diffractometer equipped with a CCD detector (radiation MoKalpha). The software DENZO-SMN was used for refinement of the unit cell and data reduction.

The SEM imaging was performed by the MERLIN Zeiss SEM field emission gun (FEG) instrument at an accelerating voltage of 5 kV using a secondary electron detector.
AFM imaging was carried out using a Park Scanning Probe Microscope (PSIA) in non contact mode. The image acquisition was performed in air at room temperature.

\begin{figure}[hb]
	\centering
	\includegraphics[width=0.48\textwidth]{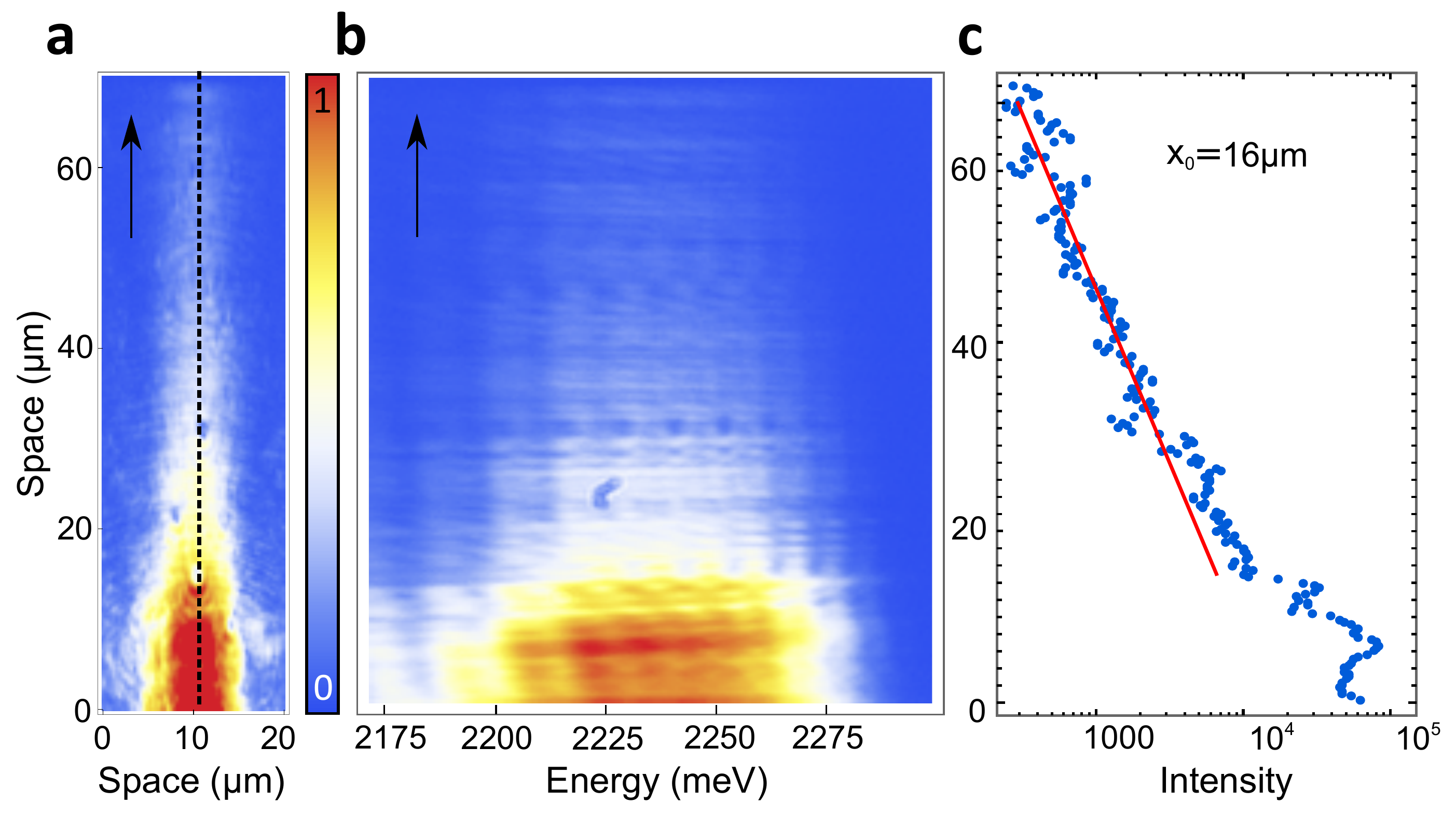}
	\caption{a) Real space propagation for a PEAI crystal of 6um obtained in a resonant configuration with a fs pulsed laser. The laser resonantly excites the Fabry-Pérot modes. b) Energy-resolved propagation map (along the black line in~S8a); it's possible to recognize the excited propagating modes outside the excitation spot. c) Intensity profile and propagation constant value, exctracted at 2230meV; the value is around 16um and is almost the same for each propagating energy.
}\label{fig:2}
\end{figure}

\noindent{\bf Optical measurements}
A home built microscope, equipped with two objectives for reflection and transmission measurements, is used to perform all optical measurements (see Fig. 1c of Ref. ~\cite{LSALerario}).
Photoluminescence (PL) is excited through a 10X objective (Rolyn-Rau, N.A.=0.3) and the PL signal is collected by A 60X oil immersion microscope objective (Olympus, N.A.=1.49). The detected signal is focused, by using a 50cm lens, into a 300mm spectrometer (Princeton Instruments, Acton Spectra Pro SP-2300) coupled to a charge coupled device (Princeton Instruments, Pixies 400). The spectrometer is equipped with two gratings, 300 g/mm and 1200 g/mm, both of them blazed at 500nm. The 300g/mm grating is used for reflectivity and PL measurements (overall spectral resolution of 2nm). A 20cm lens is additionally used  to image the objective’s back focal plane (BFP) onto the entrance slits of the spectrometer. 
Energy vs in-plane momentum reflectivity measurements (Fig. 2, Fig. 10, and Fig. 11a) are performed by using a Xenon light source (Korea Spectral Products-ASB-XE-175) that resonantly excites the Fabry-Perot modes through the oil immersion objective. A half-wave plate (AHWP10M-600) and a linear polarizer (LPVISSE100) are placed in front of the spectrometer in order to resolve the two polarizations TE and TM. 
Propagation and polarization measurements (Fig. 4) are performed by using a 640nm CW laser diode (Coherent-BioRay).
For PL measurements, a 488nm CW laser (Spectra-Physics) is used to excite the material (Fig. 3, Fig. 11, Fig. 9 and Fig. 11b) through a 10X objective. A 500nm cut-off filter (Thorlabs-FEL0500) along the detection line cuts the residual excitation laser intensity. For resonant propagation reported in Fig. S8, a 50 fs pulsed laser (Coherent, TOPAS-Prime 10Kz) is used to resonantly excite the Fabry-Perot modes.

%



\section{The scattering matrix method}
\subsection{General method}
The optical response, namely transmission and reflection coefficients, of the perovskite layered crystals is modeled through the Scattering Matrix Method (SMM), as reported in Refs. ~\cite{Li1996,Whittaker1999,Liscidini2008}. In particular, we have implemented a Python version of the method as first devised by Lifeng Li ~\cite{Li1997new}. We hereby present a short summary of this formulation by using the same notation as in the original reference ~\cite{Li1997new}. We also introduce a straightforward generalization of the method that allows to model anisotropic materials whose principal axis is along the frame of reference, i.e. described by a diagonal dielectric tensor generically expressed as
\begin{equation} \label{eq:die_ten}
\mathbf{\varepsilon} = 
\left( \begin{array}{ccc}
\varepsilon_{x} & 0 & 0 \\
0 & \varepsilon_{y} & 0 \\
0 & 0 & \varepsilon_{z} \\
\end{array} \right) \, .
\end{equation}
With reference to the original notation in Ref.~\cite{Li1997new}, the only modification needed when assuming a rectangular lattice ($\chi=0$) is to take the right dielectric tensor element, which is essential to calculate the matrices composing the eigenvalue problem that has to be solved to get the propagation modes inside the single layer. In particular, the matrix $\llbracket \varepsilon \rrbracket$ is calculated with the $\varepsilon_z$ element, the matrix $\lfloor \lceil \varepsilon \rceil \rfloor$ is calculated with $\varepsilon_x$, and $\lceil \lfloor \varepsilon \rfloor \rceil$ with $\varepsilon_y$, respectively.
All the remaining parts of the code, from creation of interface and propagation scattering matrices to the recursion algorithm, are exactly implemented as in the original reference work. 
Applying the SMM to the present situation, a further simplifying assumption can be made since the structures to be modeled are homogeneous in the $xy$ plane (the perovskite layers are not patterned). Hence, the code can actually be run by only retaining one term in the Fourier expansion of the modes, and actually working with $4 \times 4$ scattering matrices. However, we stress that the SMM presented here is general and could be applied to patterned structures.

\subsection{Simulation details.}
\begin{figure} [h]
\centering
\includegraphics[width=0.4\textwidth]{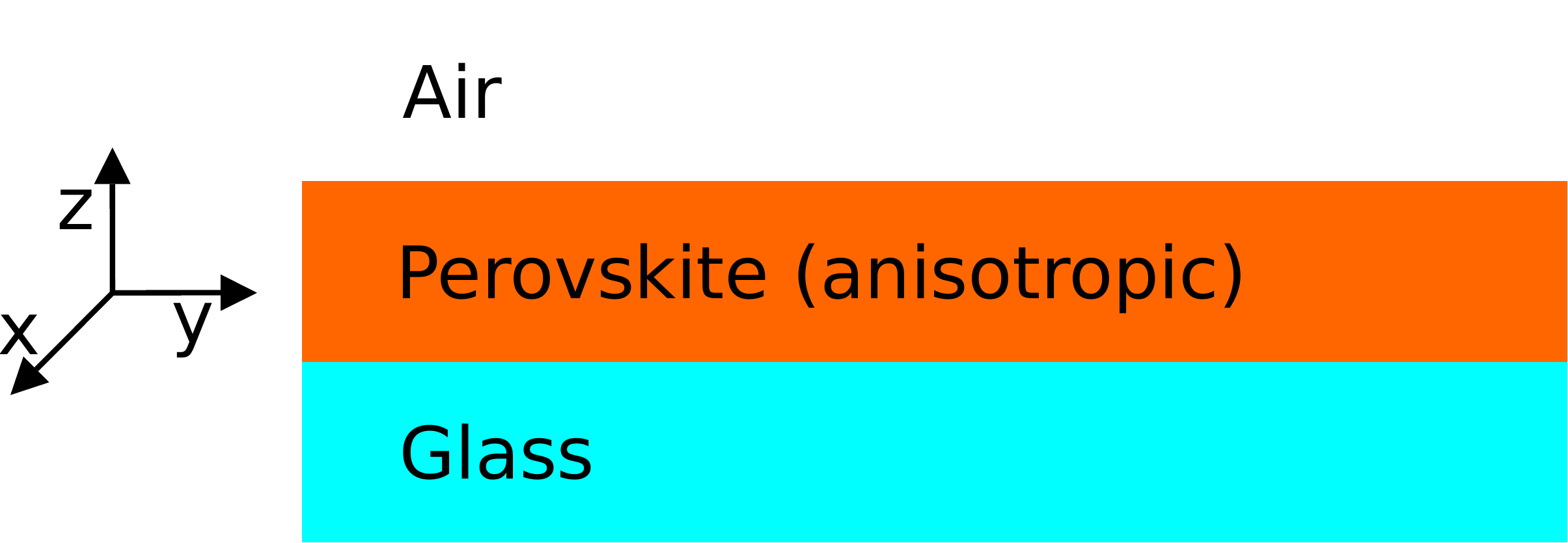}
\caption{Schematic setup of the simulation and the related reference frame.}
\label{fig:setup}
\end{figure}

The structure under study is presented in Fig. \ref{fig:setup}. Glass and air are assumed as dispersionless and isotropic media with refractive indices 1.49 and 1, respectively. The light is assumed coming from the underside of the structure, with and angle of incidence $\theta$ with respect to the normal direction. The plane of incidence is assumed to be the $yz$ plane, so we will speak of TE or $s$-polarized light when the electric field is along the $x$ direction, and of TM or $p$-polarized light when the electric field is in the $yz$ plane. 

In fact, the interest and originality of the present work mostly lies in the synthesis and optical characterization of layered  perovskite crystalline materials, composed by a periodic repetition of thin optically active layers with interposed organic barriers. In this respect, the system is very similar to a multi-quantum well structure typically grown with semiconductor materials. At difference with ordinary quantum well structures, though,  these artificial perovskites exhibit strong anisotropic behavior, in addition to an absorption peak due to excitonic-like response. We may assume that the excitonic states are confined in the inorganic layers. 

As a consequence, we  modeled these perovskites  as thin layers of anisotropic materials possessing both in-plane and out-of-plane components, namely with a dielectric tensor generically expressed as
\begin{equation} \label{eq:per_ten}
\mathbf{\varepsilon} = \left( \begin{array}{ccc}
\varepsilon_\parallel & 0 & 0 \\
0 & \varepsilon_\parallel & 0 \\
0 & 0 & \varepsilon_\perp \\
\end{array} \right) \, .
\end{equation}
To account for the anisotropic excitonic-like behavior, we assumed dispersive Lorentz expressions for both $\varepsilon_\parallel$ and $\varepsilon_\perp$, superimposed to a Cauchy background, i.e.
\begin{equation} \label{eq:epsilon}
\varepsilon(E)=\varepsilon_c(E) + \varepsilon_l(E) \, ,
\end{equation}
where 
\begin{equation} \label{eq:cauchy}
\varepsilon_c(E) = \left( n_b  +AE^2\right)^2
\end{equation}
is the general expression for the Cauchy contribution, in which $n_b$ and $A$ are parameters extracted from ellipsometry measurements in the region outside the excitonic absorption, while the Lorentz term is assumed in the usual semiconductor-like form~\cite{Andreani1994}
\begin{equation} \label{eq:lorentz}
\varepsilon_l(E)= \varepsilon_{\infty} \left[ 1+ \frac{E_{\mathrm{LT}}}{E_0-E-i\Gamma} \right] 
\end{equation}
in which $E_0$ and $\Gamma$ are the resonance energy and broadening (full width at half maximum), respectively, as extracted from the ellipsometry measurements in the region corresponding to the excitonic absorption peak. In the Lorentz expression, $E_{\mathrm{LT}}=\hbar^2 e^2 f_{\mathrm{osc}}/(2\varepsilon_0 \varepsilon_{\infty} m_0 E_0 d)$, in which $\varepsilon_0$ (vacuum dielectric permittivity), $e$ (electric charge) and $m_0$ (rest electron mass) are fundamental constants, while $f_{\mathrm{osc}}$ is the oscillator strength associated to the excitonic transition for a perovskite layer thickness $d$. 
Since in the spectral region of interest is difficult to distinguish between in-plane and out-of-plane contributions to $\varepsilon_{\infty}$ directly from ellipsometry measurements, we assumed the value extracted in the absorptive region as a mean value, and   $E_0$ and $\Gamma$ are taken to the same values for both components of the dielectric tensor. The values assumed for each of the crystalline compounds analyzed in this work are explicitly reported in table \ref{tab:par}. 
We notice that the oscillator strength is the main quantity to be adjusted and extracted through our SMM simulations by comparison with experimental reflectivity spectra, assuming different values for  in and out-of plane components, respectively. 
We remind that all energies are expressed in eV in our numerical simulations, which are used  to calculate reflectivity spectra for both TE and TM polarized input radiation, respectively. The spectra thus obtained are directly compared to the experimental ones, which allows to get a reliable quantitative estimate for the in and out-of plane values of the oscillator strengths. The results of such an analysis are reported in the main text.

\begin{table}
\centering
\begin{tabular}{ccccccc}
\hline
       & \multicolumn{2}{c}{In-plane Cauchy} & \multicolumn{2}{c}{Out-of-plane Cauchy} & \multicolumn{2}{c}{Lorentz} \\ 
Material  & $n_b$ & A & $n_b$ & A & $E_0$ & $\Gamma$  \\
\hline
BAI  & 1.628 & 0.04278 & 1.789 & 0.04386 & 2.41 & 0.0524 \\
PEAI & 1.765 & 0.04484 & 1.855 & 0.04484 & 2.39 & 0.0334 \\
OCT  & 1.490 & 0.02179 & 1.692 & 0.02179 & 2.41 & 0.0448 \\
\hline
\end{tabular}
\caption{Table with the data extracted from ellipsometry measurements for the three material synthesized and characterized in this work.}
\label{tab:par}
\end{table}


\section{Ordinary and extraordinary refractive index}

\begin{figure}[hb]
	\centering
	\includegraphics[width=0.48\textwidth]{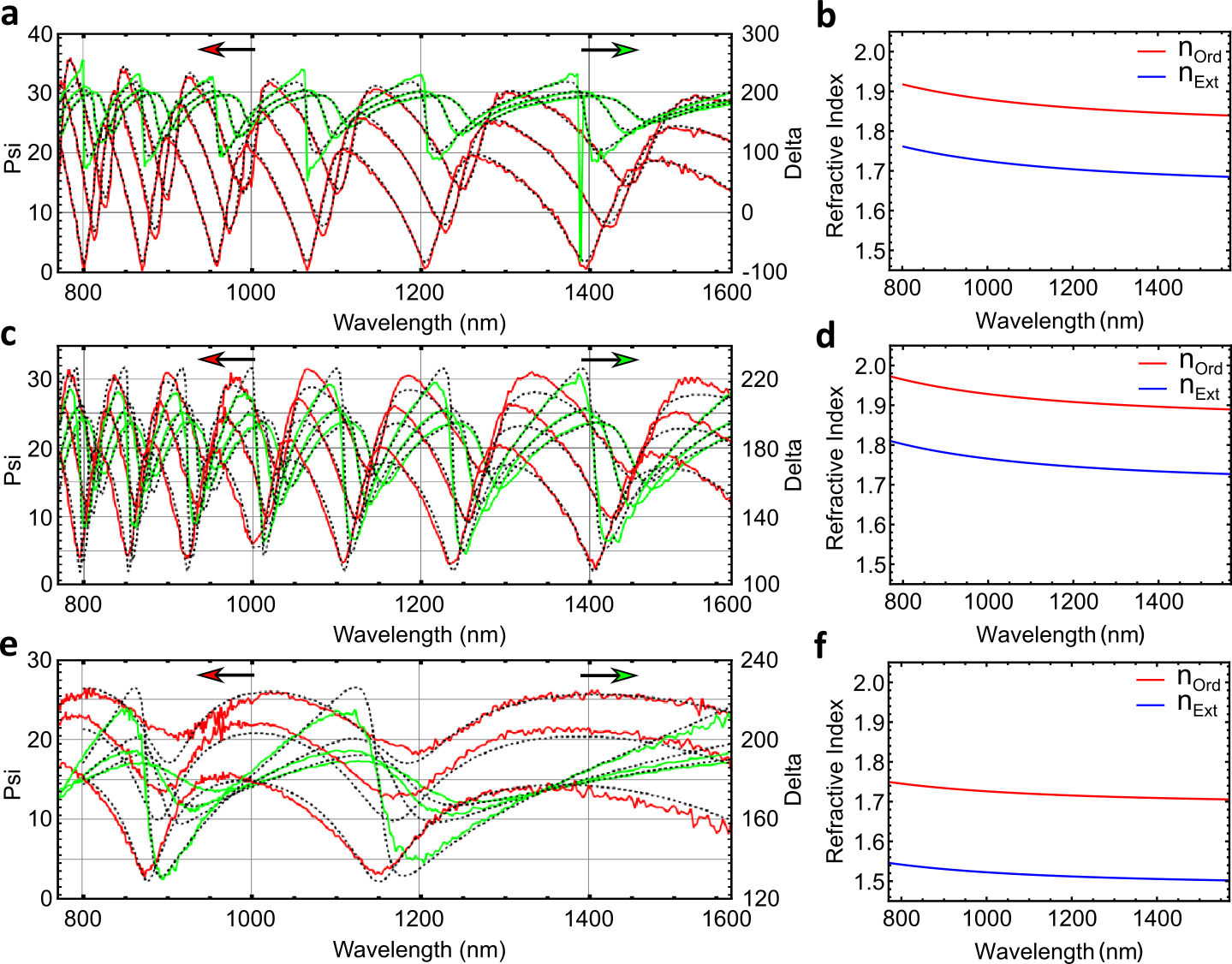}
	\caption{Fitting ellipsometric measurements in the transparent region of BAI (a), PEAI (c) and OCT (e). The corresponding ordinary and extra-ordinary refractive index are shown in b), d) and e) for BAI, PEAI and OCT, respectively.
}
\end{figure}

The real part of the extraordinary n\textsubscript{ext} and of the ordinary n\textsubscript{ord} refractive indexes have been evaluated by ellipsometric measurements (J.A. Wollam-EC-400). The single crystal perovskite is deposited on a glass substrate and the thickness is measured by a profilometer. In order to collect signal from the central part of the flake a lens system is used to reduce the spot size.  An scan over the angle of incidence is performed in the region around the Brewster angle. The experimental results for $\Psi$ and $\Delta$ are reported in the left part of Fig.~14, together with the model, for three different incident angles ($40^\circ$, $50^\circ$, $55^\circ$). 
The asymmetric oscillations in the transparent region suggest an optical anisotropy that is evident for each kind of material. We used a Cauchy model obtaining good fitting for the three materials, with the corresponding results for n\textsubscript{ext} and n\textsubscript{ord} reported in the right panel of Fig.~14.

%


%

\end{document}